\documentclass[aps,a4paper,preprint]{revtex4}%
\usepackage{amsfonts}
\usepackage{amsmath}
\usepackage{amssymb}
\usepackage{graphicx}
\usepackage{setspace}%
\providecommand{\U}[1]{\protect\rule{.1in}{.1in}}
\begin{document}
\title{Continuity and discontinuity of kirigami's high-extensibility transition: a
statistical-physics viewpoint}
\author{Midori Isobe and Ko Okumura}
\affiliation{Physics Department, Faculty of Science, Ochanomizu University}
\date{\today}

\begin{abstract}
Recently, kirigami's high extensibility has been understood as a transition in
the force-elongation curve. In this paper, we consider a model, which modifies
our previous model, to show a striking analogy between the present theory and
Landau theory of continuous thermodynamic transitions, if we regard a rotation
angle and elongation of kirigami as the order parameter and the inverse
temperature, respectively. The present study opens a new avenue in physics,
pointing out the importance of the distinction between discontinuity and
continuity of the high-extensibility transition in an elementary kirigami
structure, and showing that the mechanical response of kirigami can be
understood using the tools of statistical physics, which have been proved to
be useful in many fields of physics.

\end{abstract}
\maketitle

Origami and kirigami, Japanese traditional craft technique based on folding
and/or cutting paper, have been received extensive attention in scientific
fields because of their potential to impart sheet materials to mechanical and
functional properties with simple patterning \cite{miura1985,xu2017origami}.
Resulting applications have been frequently regarded as mechanical
metamaterials \cite{shan2015design,bertoldiexploiting} or tunable mechanical
devices \cite{bertoldi2017flexible}. The basic kirigami structure, formed by
patterning parallel cuts on a sheet, makes sheet materials highly stretchable,
which is shown even for graphene sheets
\cite{GraphenKirigami2015Nature,GraphenKirigami2014PRB}. The high
stretchability emerges from the transition from the in-plane to out-of-plane
deformation, which is accompanied by a buckling-induced rotation of each unit
of the structure \cite{isobe2016initial}. While other cut patterns have been
studied to explore versatile possibilities of the application of kirigami
\cite{rafsanjani2017buckling,hwang2018tunable}, high stretchability of
kirigami has been one of the important properties of kirigami. This property
has been applied to varieties of materials such as conducting nanocomposites
\cite{Kirigami2015NatMat}, piezoelectric materials \cite{hu2018stretchable},
metallic glass \cite{chen2018highly}, and thermally responsive materials
\cite{tang2017programmable}, and to specific devices such as stretchable
strain sensor \cite{sun2018kirigami} and flexible film bioprobe
\cite{morikawa2018ultrastretchable}. Some researchers have focused on other
available characteristics of kirigami structures. The buckling-induced
rotation has been exploited for developing solar-tracking batteries
\cite{KirigamiSolarNC2015} and dynamic shading systems \cite{yi2018developing}%
. Frictional and interfacial properties have been utilized for fabricating
soft actuator \cite{rafsanjani2018kirigami} and enhancing film adhesion
\cite{zhao2018kirigami}, respectively. Although widely studied from
application-oriented perspectives, the basic physical understanding of the
high extensibility of kirigami is still premature, which is the focus of the
present study.

\begin{figure}[h]
\includegraphics[width=0.9\textwidth]{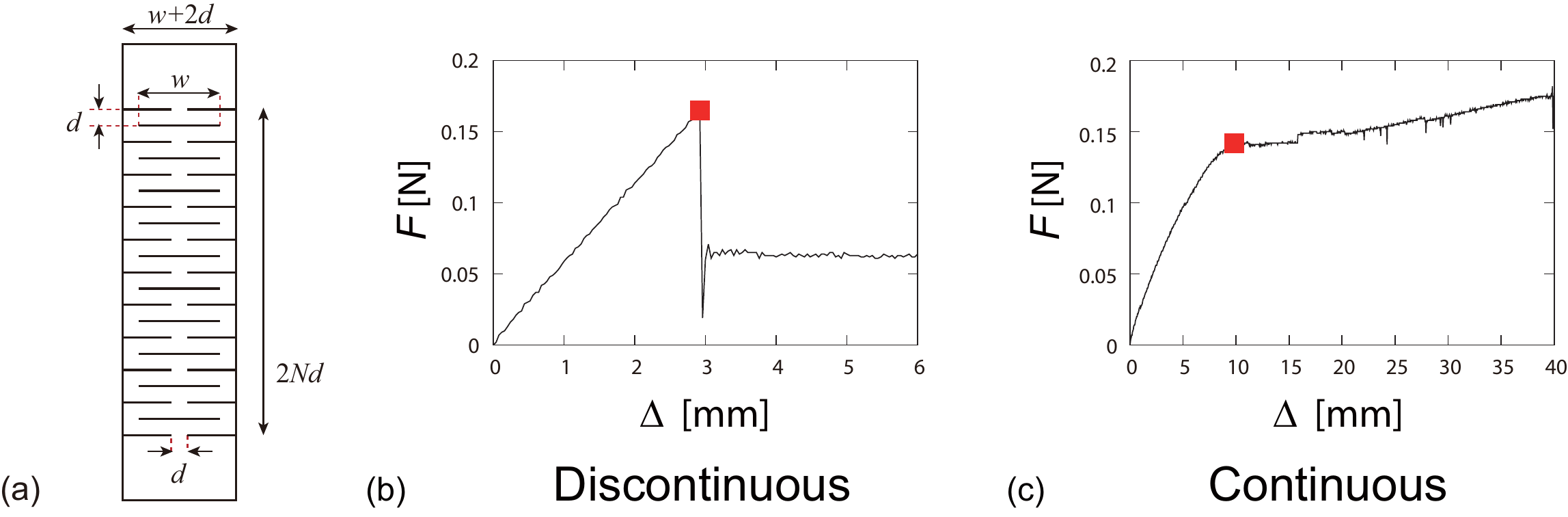}.\caption{(a) Geometry of a
simplified kirigami specimen. (b, c) Two typical examples of the force $F$ as
a function of the elongation $\Delta$ in the vicinity of transition point, one
exhibiting an abrupt (discontinuous) transition and the other exhibiting a
(piece-wise) continuous transition. The former is obtained from a ridged
polystyrene plate of Young's modulus $E=3.1$ GPa ($h=0.2$ mm, $w=40$ mm, $d=2$
mm), while the latter from a soft elastomer sheet of $E=7.9$ MPa ($h=1$ mm,
$w=30$ mm, $d=5$ mm).}%
\label{f1}%
\end{figure}

As already mentioned above, the high stretchability of the basic kirigami
structure (see Fig. \ref{f1}(a)) has been explained by a simple model based on
bending energy \cite{isobe2016initial}. Experimentally, this transition
manifests as a transition in the force-elongation curve. The transition could
be an abrupt or a smooth transition and this difference seems to be dependent
on materials used for fabricating specimens and the geometries of the cut
patterns, as shown in the results reported from various groups
\cite{chen2018highly,sun2018kirigami,hu2018stretchable} (see Discussion): we
provide typical experimental results for the two opposite cases in Fig.
\ref{f1}(b) and (c). An interesting issue here is whether this transition
could be regarded as a thermodynamic transition and if this is the case
whether the high-extensibility transition is continuous or discontinuous. In
fact, in our recent work \cite{isobe2019discontinuity}, we showed that our
previous model proposed in \cite{isobe2016initial} predicts a discontinuous
transition and the prediction on the ratio between the forces just before and
after the jump agree semi-quantitatively with experimental data obtained from
kirigami samples made of\ Kent paper.

Here, we generalize our previous model and show that the kirigami's
high-extensibility transition can physically be identified with Landau theory
of the second-order transition \cite{cardy1996scaling,goldenfeld2018lectures},
if we regard a rotation angle $\theta$ and elongation $\delta$ of each unit as
the order parameter and the inverse temperature, respectively. We briefly
discuss the possible mechanism of the kirigami's transition becoming a
discontinuous transition.

\noindent\textit{Geometry of a unit of kirigami under tension.---} In this
study, we consider a simplified kirigami structure, whose non-deformed
geometry is specified in Fig. \ref{f1}(a). An elementary unit of kirigami is
defined as each of $2N$ strips, whose width, height, and thickness are
respectively given by $w,d$, and $h$ with $w\gg d$ (see Fig. \ref{f1}(a)). The
($2n-1$)-th elementary unit (from the top) is connected at the left and right
edges of volume $hd^{2}$ with the $2n$-th elementary unit ($n=1,\cdots N$).

\begin{figure}[h]
\includegraphics[width=0.8\textwidth]{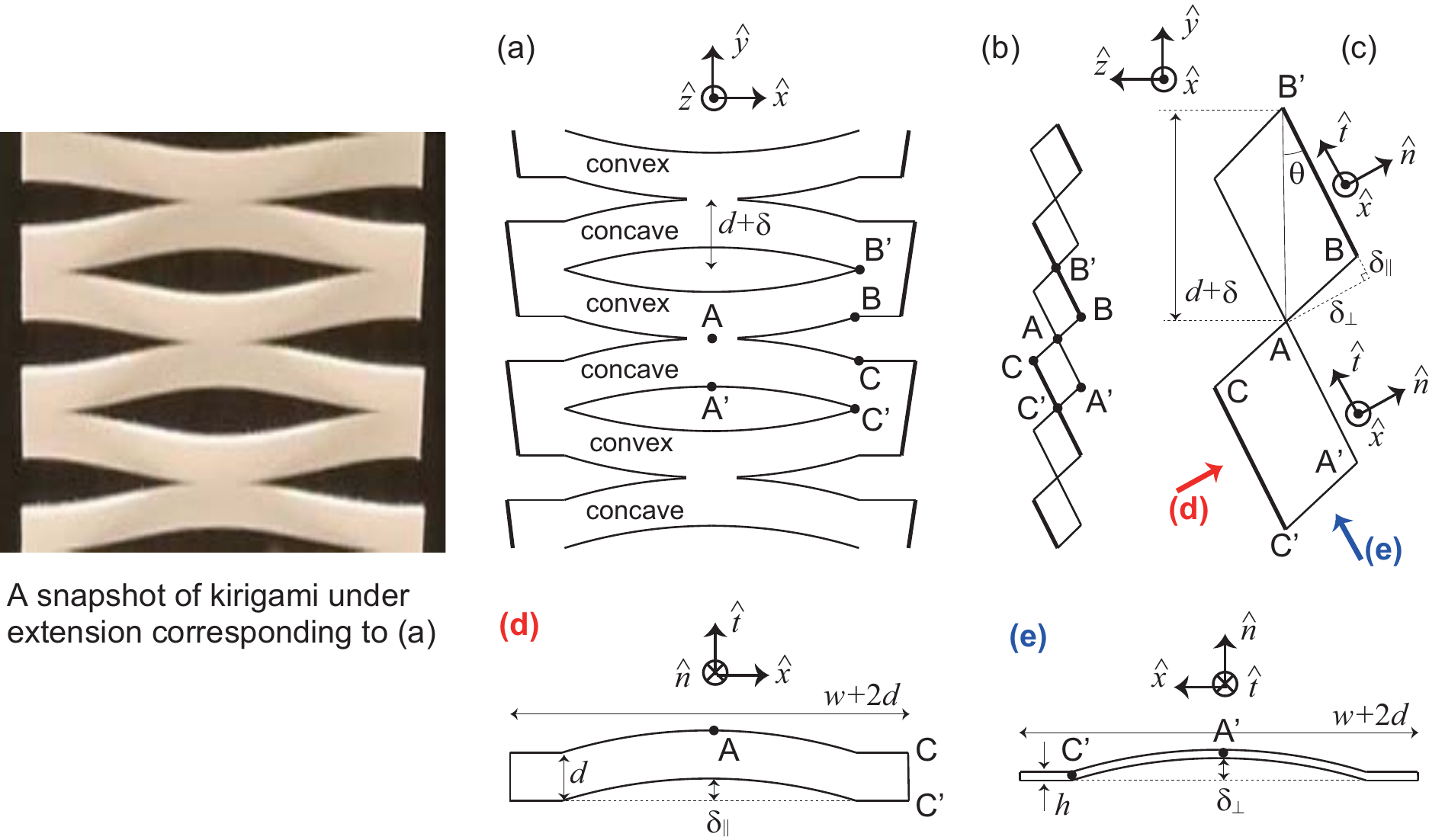}\caption{Kirigami's geometry
under a given elongation $\delta=\Delta/(2N)$. In the figure, the vectors,
$\hat{x},\hat{y},\hat{t},$ and $\hat{n}$ are unit vectors, defined in the
text, and used to show the plane on which the corresponding illustration is
drawn. (a) Front view of an extended kirigami. (b) Side view. (c) Magnified
side view. (d) The unit of kirigami containing the points A, A', C, and C'
seen from the direction $\hat{n}$. (e) The same unit seen from the direction
of $\hat{t}$.}%
\label{f2}%
\end{figure}

In the previous study, we considered two modes of deformation. In the in-plane
deformation, the central part of volume $hwd$ of the ($2n-1$)-th [$2n$-th]
elementary unit bends "downwards" ["upwards"] such that the arc of length $w$
and the center of this arc are on the original plane and the center is located
below [above] the unit. In the out-of-plane deformation, the central part of
volume $hwd$ of the ($2n-1$)-th [$2n$-th] elementary unit bends "forwards"
["backwards"] such that the arc of length $w$ and the center of this arc are
on the plane rotated from the original plane by a finite (small) angle and the
center is located in front of [behind] the unit.

In the present study, we allow the simultaneous existence of these two modes.
(We disallow it in the previous study.) Such a general deformation is
described in Fig. \ref{f2}, which is explained in detail below. The in-plane
and out-of-plane deformation in the previous study correspond to the special
cases in which $(\delta_{\perp},\theta)=(0,0)$ and in which $\delta_{\Vert}%
=0$, respectively.

In Fig. \ref{f2}, the specimen is elongated by the amount $\Delta$ and, thus,
each unit is elongated by $\delta=\Delta/(2N)$, as shown in Fig. \ref{f2}(a)
and the corresponding snapshot on the left side. Here, the unit vectors in the
$x,y$ and $z$ directions are given by $\hat{x}$, $\hat{y}$, and $\hat{z}$,
respectively. In Fig. \ref{f2}(a), drawn on the $x-y$ plane, the surface of
the central part of volume $hwd$ of the ($2n-1$)-th [$2n$-th] unit is
"concave" ["convex"] such that, for example, C and C' are closer to you than A
and A' [A is closer to you than B]. The side view of Fig. \ref{f2} (a) is
given in (b), which is drawn on the $y-z$ plane. A part of (b) is magnified in
(c), in which the unit vectors $\hat{t}$ and $\hat{n}$ are respectively shown
as the tangential and normal vectors on the surface of an above-mentioned
element of kirigami, with the angle between the vector $\hat{t}$ and $\hat{y}$
being $\theta$. (Although each element is not on a plane but on a curved
surface, $\hat{t}$ is identical at any point on the curved unit surface (and
thus uniquely defined), and $\hat{n}$ can also be uniquely defined) In
general, for each element "rotated by the angle $\theta$" under the given
elongation $\delta$, the deformation is characterized by the vector
\begin{equation}
\vec{\delta}_{T}=\delta_{\Vert}\hat{t}+\delta_{\perp}\hat{n}. \label{eqdel}%
\end{equation}
In Fig. \ref{f2}(c), vectors $\overrightarrow{AB}$, $\overrightarrow{CA}$, and
$\overrightarrow{C^{\prime}A^{\prime}}$ are, for example, identical to the
vector $\vec{\delta}_{T}$. As clear from Fig. \ref{f2}(c), $\delta_{\Vert}$
and $\delta_{\perp}$ satisfy the following relations:
\begin{align}
(d+\delta)\cos\theta &  =d+\delta_{\Vert}\label{eq1}\\
(d+\delta)\sin\theta &  =\delta_{\perp} \label{eq2}%
\end{align}
This means that as illustrated in (d) and (e), the deformation of each element
can be regarded as a superposition of the bending in the $\hat{t}-x$ plane
(the arc and its center characterizing the bending are on the $\hat{t}-x$
plane; see (d)) and the bending in the direction normal to this plane (the arc
and its center are on $\hat{n}-x$ plane; see (e)), which will be called
"in-plane" and "out-of-plane" deformations, respectively. Note that "the
plane" here does not refer to the original $x-y$ plane but the $\hat{t}-x$ plane.

\noindent\textit{Deformation energy of a unit of kirigami.---} From the
observations we have seen in Fig. \ref{f2}(d) and (e), according to the
standard formula for the bending energy, the energy for the specified
deformation per unit element is given for $w\gg d,\delta$ by
\begin{align}
U(\delta,\theta)  &  =U_{\parallel}+U_{\perp}\label{U}\\
U_{\parallel}  &  =kE\frac{hd^{3}}{w^{3}}\delta_{\Vert}^{2}\label{U1}\\
U_{\perp}  &  =kE\frac{h^{3}d}{w^{3}}\delta_{\perp}^{2} \label{U2}%
\end{align}
where $E$ is Young's modulus \cite{Landau,isobe2016initial} (see Sec. A of
\cite{SM1}). In fact, the numerical coefficient $k$ depends on the boundary
conditions for bending and $k=(8/3)/(1-\nu^{2})$ with $\nu$ Poisson's ratio
\cite{Landau} if we consider that the original straight line of length $w$
becomes a part of an arc assuming that $w\gg d,h$ as in the illustrations in
Fig. \ref{f2}. Other boundary conditions tend to increase the bending
energies. (We have not considered "net" stretching energy, which can be
justified in the present case; see Sec. A of \cite{SM1} for the details.)

The independent variables for the above energy for a given set of $w$, $d$,
and $h$ are in fact $\delta$ and $\theta$, as seen from Eqs. (\ref{eq1}) and
(\ref{eq2}). By renormalizing the energy and lengths by using the energy unit
$kEhd^{5}/w^{3}$ and the length unit $d$, we obtain a dimensionless
expression:
\begin{align}
\tilde{U}(\delta,\theta)  &  =\tilde{U}_{\parallel}+\tilde{U}_{\perp
}\label{U12c}\\
\tilde{U}_{\parallel}  &  =((1+\tilde{\delta})\cos\theta-1)^{2}\label{U12}\\
\tilde{U}_{\perp}  &  =\tilde{h}^{2}((1+\tilde{\delta})\sin\theta)^{2}
\label{U12b}%
\end{align}
with $\tilde{\delta}=\delta/d$ and $\tilde{h}=h/d$.

\begin{figure}[h]
\includegraphics[width=0.3\textwidth]{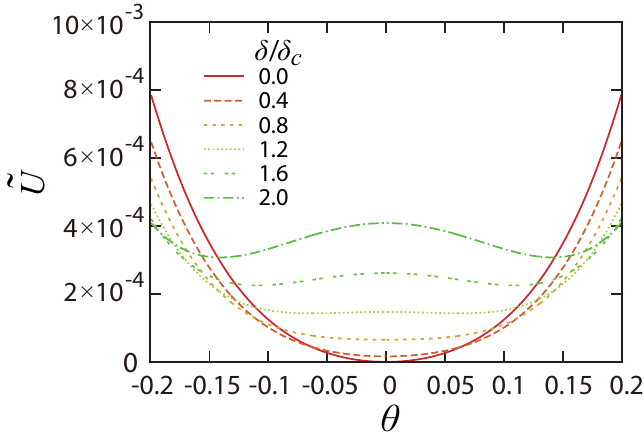}\caption{Free energy $\tilde
{U}$ as a function of $\theta$ for various $\delta$ at $h/d=0.1$.}%
\label{f3}%
\end{figure}

\noindent\textit{Transition behavior based on the energy.---} The energy
obtained above behaves as the Landau free energy for the second order
transition, if we identify $\theta$ as the order parameter and $\delta$ as the
inverse temperature. Figure \ref{f3} shows profiles of $\tilde{U}$ as a
function of $\theta$ for various $\delta$ at $h/d=0.1$; at small $\delta$ the
energy minimum appears at $\theta=0$ but as $\delta$ increases two minima
appear symmetrically at $\theta=\pm$ $\theta^{\ast}$ with a finite
$\theta^{\ast}$ ($>0$), which parallels Landau theory of critical phenomena.
This feature is generic for the analytical structures as long as $h\ll d$ is
satisfied, as discussed in Sec. B of \cite{SM1}, in which physical origins of
the emergence of Landau's scenario is elucidated. For the representation of
the plots, we have introduced $\delta_{c}$ as
\begin{equation}
\tilde{\delta}_{c}=\delta_{c}/d=\tilde{h}^{2}/(1-\tilde{h}^{2})\simeq\tilde
{h}^{2} \label{delc}%
\end{equation}

As seen in Fig. \ref{f3} (and shown precisely in Eq. (\ref{theta}) below), the
quantity $\delta_{c}$ corresponds to "the inverse critical temperature," i.e.,
the value of $\delta$ at which the energy minima start to appear at nonzero
values of the order parameter $\theta$. The corresponding scaling exponent is
$1/2$, as shown in the following analytical expression derived in Sec. B of
\cite{SM1} with the assumption $\theta\ll1$ and thus exact near the critical
point for the present model:
\begin{equation}
\theta^{\ast}=\left\{
\begin{array}
[c]{cc}%
0 & \text{for }\delta<\delta_{c}\\
\left(  2(1-\tilde{h}^{2})(\tilde{\delta}-\tilde{\delta}_{c})\right)  ^{1/2} &
\text{for }\delta>\delta_{c}%
\end{array}
\right.  \label{theta}%
\end{equation}
Here, $\theta^{\ast}$ is theoretically predicted value of the rotation angle
that is obtained as the minimum of $\tilde{U}(\tilde{\delta},\theta)$ in terms
of $\theta$, by finding one of the solutions of the following equation for
$\theta$:
\begin{equation}
\frac{\partial\tilde{U}(\tilde{\delta},\theta)}{\partial\theta}=0. \label{eqT}%
\end{equation}

The order parameter in the present theory predicts a continuous transition, as
seen in Eq. (\ref{theta}), whereas the order parameter in the previous theory
\cite{isobe2016initial,isobe2019discontinuity} predicts a discontinuous
transition. As summarized in Sec. C of \cite{SM1}, the quantity $\theta^{\ast
}$ obtained in the previous theory jumps at $\delta=2\delta_{c}$ from
$\theta=0$ to $\theta=\theta_{c}^{\ast}$ where $\theta_{c}^{\ast}$ is
$\theta^{\ast}$ at $\delta=2\delta_{c}$, i.e., $\theta_{c}^{\ast}=\tan
^{-1}\left(  4\tilde{\delta}_{c}\right)  ^{1/2}\simeq2\tilde{h}$ (see Eq.
(33)). In other words, the expressions for $\theta^{\ast}$ in the present and
previous models give the critical elongations, $\delta=\delta_{c}$ and
$\delta=2\delta_{c}$, respectively, and thus the two models are similar in
that the predicted critical elongations are the same at the level of scaling
laws but are different because of an extra factor 2.

\begin{figure}[h]
\includegraphics[width=\textwidth]{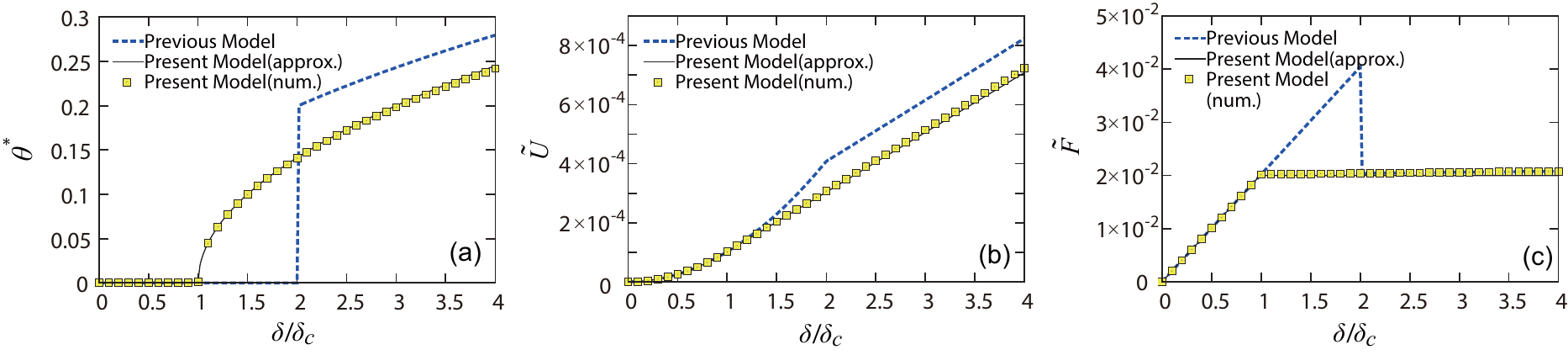}\caption{ (a) The value
$\theta^{\ast}$ of "the order parameter" $\theta$ predicted by the theories as
a function of "the inverse temperature" $\delta$ for $h/d=0.1$, demonstrating
that the present and previous models are continuous and discontinuous,
respectively. (b) Normalized free energy minimized with respect to $\theta$ as
a function of $\delta$ and (c) Normalized force - elongation curve for
$h/d=0.1$.}%
\label{f4}%
\end{figure}

The expression for $\theta^{\ast}$ in Eq. (\ref{theta}) is plotted in Fig.
\ref{f4}(a) under the label "Present Model (approx.), which confirms a
continuous transition. The plot labeled as "Present Model (num.)" in Fig.
\ref{f4}(a) is obtained numerically finding the root of the condition in Eq.
(\ref{eqT}) without using the approximation $\theta\ll1$. The numerical and
approximate plots agree well with each other, which shows that the analytical
expression is a good approximation in the range of $\delta$.

The plot labeled as "Previous Model" in Fig. \ref{f4}(a) is based on the
analytical expression obtained in the previous model (see Sec. C of
\cite{SM1}). By comparing the three plots in Fig. \ref{f4}(a), we can
re-confirm the similarities and differences in the $\delta$-dependence of
$\theta^{\ast}$ in the two models discussed above on the basis of the
analytical expression.

The analytical expressions for $\tilde{U}(\tilde{\delta},\theta^{\ast})$,
which is $\tilde{U}(\tilde{\delta},\theta)$\ evaluated at $\theta=\theta
^{\ast}$, and the force defined by $\tilde{F}=\left\vert -\partial\tilde
{U}(\tilde{\delta},\theta^{\ast})/\partial\tilde{\delta}\right\vert $ (the
force $F$ is normalized here by the unit $kEhd^{4}/w^{3}$) obtained in the
present model are summarized in Sec. B of \cite{SM1}, and predict the
following features of the present model: (1) $\tilde{U}(\tilde{\delta}%
,\theta^{\ast})$ scales with $\tilde{\delta}^{2}$ and $\tilde{\delta}$ for
$\delta<\delta_{c}$ and $\delta>\delta_{c}$, respectively, and the two
branches are matched at $\delta=\delta_{c}$. (2) $\tilde{F}(\tilde{\delta})$
scales with $\tilde{\delta}$ and $\tilde{\delta}^{0}$ for $\delta<\delta_{c}$
and $\delta>\delta_{c}$, respectively, and the two branches are matched at
$\delta=\delta_{c}$ (because $\tilde{\delta}\simeq\tilde{h}^{2}$ at
$\delta=\delta_{c}$ for $\tilde{h}\ll1$).

There are similarities and differences between the analytical expressions for
$\tilde{U}$ and $\tilde{F}$ in the present model (see Sec. B of \cite{SM1})
and those in previous model (see Sec. C \cite{SM1}). (1) In both models, the
quadratic $\tilde{\delta}$-dependence of $\tilde{U}(\tilde{\delta})$ is
switched to the linear $\tilde{\delta}$-dependence at the transition point.
(2) The spring constant (slope of the force-elongation curve) before the
transition is exactly the same in the two theories. However, there are
important qualitative differences in the two theories. (1) In the previous
theory $\tilde{U}(\tilde{\delta})$ is piecewise continuous (discontinuous at
the transition $\delta=2\delta_{c}$), whereas it is continuous at the
transition $\delta=\delta_{c}$ in the present theory. (2) Accordingly, the
force-elongation curve shows a discontinuous jump at the transition point in
the previous theory, whereas it is (piecewise)\ continuous in the present
theory; $\tilde{F}(\tilde{\delta})$ in Eq. (35) shows a drop from $\tilde
{F}(\tilde{\delta})=4\tilde{\delta}_{c}$ to $2\tilde{\delta}_{c}$ at
$\delta=2\delta_{c}$, while $\tilde{F}(\tilde{\delta})$ in Eq. (32) shows a
jump-less crossover ($\tilde{F}(\tilde{\delta})=2\tilde{\delta}_{c}$ at
$\delta=\delta_{c}\pm\varepsilon$ with $\varepsilon$ a small number).

The analytical expressions for $\tilde{U}$ and $\tilde{F}$ in the present
model are plotted in Fig. \ref{f4}(b) and (c) under the label "Present Model
(approx.)," from which we can re-confirmed the above features. The plots
labeled "Present Model (num.)" in Fig. \ref{f4}(b) and (c) are, respectively,
given by numerically evaluating Eq. (\ref{U12c}) at $\theta=\theta^{\ast}$
(here, the value numerically obtained, by solving Eq. (\ref{eqT})) and by
differentiating the plot in (b) with respect to $\delta$, on the basis of the
definition of $\tilde{F}$ given above. We see that the plots labeled "Present
Model (approx.)" and "Present Model (num.)" in (b) agree with each other, as
in (c), which justifies again our approximate analytical expressions.

The plots for the previous model, labeled as "Previous Model," Fig.
\ref{f4}(b) and (c) are based on analytical expressions summarized in Sec. C
of \cite{SM1}. By comparing the three plots in (b), as well as in (c), we can
re-confirm the above-discussed similarities and differences. Note that Fig.
\ref{f4}(c) corresponds to the experimental results shown in Fig. \ref{f1}(b)
and (c).

\noindent\textit{Discussion.---} The previous \cite{isobe2016initial} and
present theories predict the same scaling laws for the spring constant in the
first regime, the critical elongation, and the force at large elongation.
Thus, the agreement of these predictions with our experimental data shown in
\cite{isobe2016initial,isobe2019discontinuity} justifies the present theory,
as well as the previous theory.

One of the important issues demonstrated in this paper is the continuity and
discontinuity in the high-extensiblity transition of kirigami for the present
simple slit geometry. As shown above, this distinction is captured clearly in
the framework of statistical physics if we focus on the appropriate order
parameter, the rotation angle, and identify the elongation with the inverse temperature.

According to several previous studies, the transition in force-elongation
curve seems to be discontinuous (e.g., \cite{hu2018stretchable}) or continuous
(e.g., \cite{chen2018highly,sun2018kirigami}) depending on conditions,
although there has not been any systematic experimental investigation on the
topic, except for a very recent article (but mainly in a more complex
arrangement of silts) \cite{rafsanjani2019propagation}. As demonstrated in
Fig. \ref{f1}(b) and (c), it has been suggested that when kirigami is made of
paper or ridged plastic sheets transitions tend to be discontinuous
(corresponding to our previous model, in which the simultaneous existence of
"purely in-plane and out-of-plane deformations" is disallowed), and when
kirigami is made of soft gels transitions tend to be continuous (corresponding
to our present model, in which the simultaneous existence is allowed). We note
here that, by developing the technique of elastic charges, the
force-elongation curve has recently been discussed in
\cite{moshe2019nonlinear}, and the importance of stress relief is discussed in
\cite{moshe2019kirigami} focusing only on the continuous case for a rather
different non-slit geometry.

To deepen our understanding of the distinction between the continuity and
discontinuity, we have to explore the effects ignored in our theories, such as
friction and plastic deformation, and perform experiments focusing on these
aspects with systematically changing parameters, such as thickness and elastic
modulus. These topics will be discussed elsewhere.

In the emerging field of mechanics of metamaterials, the connection with
critical phenomena in statistical physics has been lacking, although the
scaling law (the usefulness of which is for physicists deeply rooted in the
lessons learned from critical phenomena) and thermodynamic concepts have been
explored in a number of recent publications
\cite{yang2018multistable,rafsanjani2019propagation} and connections to
bifurcation and nonlinear physics have been stressed in the literature,
probably from historical reasons
\cite{pippard1985response,audoly2010elasticity,cedolin2010stability}. The key
words such as the critical exponent and the order parameter, which are
indispensable to describe critical phenomena, have been absent in the
literature (except for a few \cite{bobnar2011euler}), although the
classic Euler buckling can be discussed in the framework of Landau theory of
critical phenomena, as explicitly demonstrated in see Sec. D of \cite{SM1}. (A
scaling relation similar to Eq. (\ref{theta}) is shown for kirigami actuators
\cite{dias2017kirigami}, which is again not connected to critical phenomena.)
Although the present kirigami's transition is not a critical phenomenon, the
remarkable analogy to Landau theory of critical phenomena demonstrated in the
present study points out to the researchers in the field the potential of the
powerful tools of statistical physics, which have been proved to be useful in
many fields of physics including nonequilibrium statistical physics
\cite{livi2017nonequilibrium}.

\begin{acknowledgments}
This work was partly supported by Grant-in-Aid for Scientific Research (A)
(No. 24244066) of JSPS, Japan, and by ImPACT Program of Council for Science,
Technology and Innovation (Cabinet Office, Government of Japan; No:
2014-PM01-02-01). M.I. is supported by the Japan Society for the Promotion of
Science Research Fellowships for Young Scientists (No. 17J04315). The authors
thank Professor Edward Foley (Ochanomizu University) for valuable comments on English.
\end{acknowledgments}



\begin{thebibliography}{10}
\expandafter\ifx\csname url\endcsname\relax
  \def\url#1{\texttt{#1}}\fi
\expandafter\ifx\csname urlprefix\endcsname\relax\def\urlprefix{URL }\fi
\providecommand{\bibinfo}[2]{#2}
\providecommand{\eprint}[2][]{\url{#2}}

\bibitem{miura1985}
\bibinfo{author}{Miura, K.}
\newblock \bibinfo{title}{Method of packaging and deployment of large membranes
  in space}.
\newblock \emph{\bibinfo{journal}{The Institute of Space and Astronautical
  Science Report}} \textbf{\bibinfo{volume}{618}}, \bibinfo{pages}{1}
  (\bibinfo{year}{1985}).

\bibitem{xu2017origami}
\bibinfo{author}{Xu, L.}, \bibinfo{author}{Shyu, T.~C.} \&
  \bibinfo{author}{Kotov, N.~A.}
\newblock \bibinfo{title}{Origami and kirigami nanocomposites}.
\newblock \emph{\bibinfo{journal}{ACS nano}} \textbf{\bibinfo{volume}{11}},
  \bibinfo{pages}{7587--7599} (\bibinfo{year}{2017}).

\bibitem{shan2015design}
\bibinfo{author}{Shan, S.}, \bibinfo{author}{Kang, S.~H.},
  \bibinfo{author}{Zhao, Z.}, \bibinfo{author}{Fang, L.} \&
  \bibinfo{author}{Bertoldi, K.}
\newblock \bibinfo{title}{Design of planar isotropic negative poisson's ratio
  structures}.
\newblock \emph{\bibinfo{journal}{Extreme Mechanics Letters}}
  \textbf{\bibinfo{volume}{4}}, \bibinfo{pages}{96--102}
  (\bibinfo{year}{2015}).

\bibitem{bertoldiexploiting}
\bibinfo{author}{Kochmann, D.~M.} \& \bibinfo{author}{Bertoldi, K.}
\newblock \bibinfo{title}{Exploiting microstructural instabilities in solids
  and structures: From metamaterials to structural transitions}.
\newblock \emph{\bibinfo{journal}{Applied mechanics reviews}}
  \textbf{\bibinfo{volume}{69}}, \bibinfo{pages}{050801}
  (\bibinfo{year}{2017}).

\bibitem{bertoldi2017flexible}
\bibinfo{author}{Bertoldi, K.}, \bibinfo{author}{Vitelli, V.},
  \bibinfo{author}{Christensen, J.} \& \bibinfo{author}{van Hecke, M.}
\newblock \bibinfo{title}{Flexible mechanical metamaterials}.
\newblock \emph{\bibinfo{journal}{Nature Reviews Materials}}
  \textbf{\bibinfo{volume}{2}}, \bibinfo{pages}{17066} (\bibinfo{year}{2017}).

\bibitem{GraphenKirigami2015Nature}
\bibinfo{author}{Blees, M.~K.} \emph{et~al.}
\newblock \bibinfo{title}{Graphene kirigami}.
\newblock \emph{\bibinfo{journal}{Nature}} \textbf{\bibinfo{volume}{524}},
  \bibinfo{pages}{204--207} (\bibinfo{year}{2015}).

\bibitem{GraphenKirigami2014PRB}
\bibinfo{author}{Qi, Z.}, \bibinfo{author}{Campbell, D.~K.} \&
  \bibinfo{author}{Park, H.~S.}
\newblock \bibinfo{title}{Atomistic simulations of tension-induced large
  deformation and stretchability in graphene kirigami}.
\newblock \emph{\bibinfo{journal}{Phys. Rev. B}} \textbf{\bibinfo{volume}{90}},
  \bibinfo{pages}{245437} (\bibinfo{year}{2014}).

\bibitem{isobe2016initial}
\bibinfo{author}{Isobe, M.} \& \bibinfo{author}{Okumura, K.}
\newblock \bibinfo{title}{Initial rigid response and softening transition of
  highly stretchable kirigami sheet materials}.
\newblock \emph{\bibinfo{journal}{Scientific reports}}
  \textbf{\bibinfo{volume}{6}} (\bibinfo{year}{2016}).

\bibitem{rafsanjani2017buckling}
\bibinfo{author}{Rafsanjani, A.} \& \bibinfo{author}{Bertoldi, K.}
\newblock \bibinfo{title}{Buckling-induced kirigami}.
\newblock \emph{\bibinfo{journal}{Physical review letters}}
  \textbf{\bibinfo{volume}{118}}, \bibinfo{pages}{084301}
  (\bibinfo{year}{2017}).

\bibitem{hwang2018tunable}
\bibinfo{author}{Hwang, D.-G.} \& \bibinfo{author}{Bartlett, M.~D.}
\newblock \bibinfo{title}{Tunable mechanical metamaterials through hybrid
  kirigami structures}.
\newblock \emph{\bibinfo{journal}{Scientific reports}}
  \textbf{\bibinfo{volume}{8}}, \bibinfo{pages}{3378} (\bibinfo{year}{2018}).

\bibitem{Kirigami2015NatMat}
\bibinfo{author}{Shyu, T.~C.} \emph{et~al.}
\newblock \bibinfo{title}{A kirigami approach to engineering elasticity in
  nanocomposites through patterned defects}.
\newblock \emph{\bibinfo{journal}{Nature Mater.}}
  \textbf{\bibinfo{volume}{14}}, \bibinfo{pages}{785--789}
  (\bibinfo{year}{2015}).

\bibitem{hu2018stretchable}
\bibinfo{author}{Hu, N.} \emph{et~al.}
\newblock \bibinfo{title}{Stretchable kirigami polyvinylidene difluoride thin
  films for energy harvesting: Design, analysis, and performance}.
\newblock \emph{\bibinfo{journal}{Physical Review Applied}}
  \textbf{\bibinfo{volume}{9}}, \bibinfo{pages}{021002} (\bibinfo{year}{2018}).

\bibitem{chen2018highly}
\bibinfo{author}{Chen, S.}, \bibinfo{author}{Chan, K.}, \bibinfo{author}{Yue,
  T.} \& \bibinfo{author}{Wu, F.}
\newblock \bibinfo{title}{Highly stretchable kirigami metallic glass structures
  with ultra-small strain energy loss}.
\newblock \emph{\bibinfo{journal}{Scripta Materialia}}
  \textbf{\bibinfo{volume}{142}}, \bibinfo{pages}{83--87}
  (\bibinfo{year}{2018}).

\bibitem{tang2017programmable}
\bibinfo{author}{Tang, Y.} \emph{et~al.}
\newblock \bibinfo{title}{Programmable kiri-kirigami metamaterials}.
\newblock \emph{\bibinfo{journal}{Advanced Materials}}
  \textbf{\bibinfo{volume}{29}}, \bibinfo{pages}{1604262}
  (\bibinfo{year}{2017}).

\bibitem{sun2018kirigami}
\bibinfo{author}{Sun, R.} \emph{et~al.}
\newblock \bibinfo{title}{Kirigami stretchable strain sensors with enhanced
  piezoelectricity induced by topological electrodes}.
\newblock \emph{\bibinfo{journal}{Applied Physics Letters}}
  \textbf{\bibinfo{volume}{112}}, \bibinfo{pages}{251904}
  (\bibinfo{year}{2018}).

\bibitem{morikawa2018ultrastretchable}
\bibinfo{author}{Morikawa, Y.} \emph{et~al.}
\newblock \bibinfo{title}{Ultrastretchable kirigami bioprobes}.
\newblock \emph{\bibinfo{journal}{Advanced healthcare materials}}
  \textbf{\bibinfo{volume}{7}}, \bibinfo{pages}{1701100}
  (\bibinfo{year}{2018}).

\bibitem{KirigamiSolarNC2015}
\bibinfo{author}{Lamoureux, A.}, \bibinfo{author}{Lee, K.},
  \bibinfo{author}{Shlian, M.}, \bibinfo{author}{Forrest, S.~R.} \&
  \bibinfo{author}{Shtein, M.}
\newblock \bibinfo{title}{Dynamic kirigami structures for integrated solar
  tracking}.
\newblock \emph{\bibinfo{journal}{Nature communications}}
  \textbf{\bibinfo{volume}{6}} (\bibinfo{year}{2015}).

\bibitem{yi2018developing}
\bibinfo{author}{Yi, Y.~K.}, \bibinfo{author}{Yin, J.} \&
  \bibinfo{author}{Tang, Y.}
\newblock \bibinfo{title}{Developing an advanced daylight model for building
  energy tool to simulate dynamic shading device}.
\newblock \emph{\bibinfo{journal}{Solar Energy}}
  \textbf{\bibinfo{volume}{163}}, \bibinfo{pages}{140--149}
  (\bibinfo{year}{2018}).

\bibitem{rafsanjani2018kirigami}
\bibinfo{author}{Rafsanjani, A.}, \bibinfo{author}{Zhang, Y.},
  \bibinfo{author}{Liu, B.}, \bibinfo{author}{Rubinstein, S.~M.} \&
  \bibinfo{author}{Bertoldi, K.}
\newblock \bibinfo{title}{Kirigami skins make a simple soft actuator crawl}.
\newblock \emph{\bibinfo{journal}{Science Robotics}}
  \textbf{\bibinfo{volume}{3}}, \bibinfo{pages}{eaar7555}
  (\bibinfo{year}{2018}).

\bibitem{zhao2018kirigami}
\bibinfo{author}{Zhao, R.}, \bibinfo{author}{Lin, S.}, \bibinfo{author}{Yuk,
  H.} \& \bibinfo{author}{Zhao, X.}
\newblock \bibinfo{title}{Kirigami enhances film adhesion}.
\newblock \emph{\bibinfo{journal}{Soft matter}} \textbf{\bibinfo{volume}{14}},
  \bibinfo{pages}{2515--2525} (\bibinfo{year}{2018}).

\bibitem{isobe2019discontinuity}
\bibinfo{author}{Isobe, M.} \& \bibinfo{author}{Okumura, K.}
\newblock \bibinfo{title}{Discontinuity in the in-plane to out-of-plane
  transition of kirigami}.
\newblock \emph{\bibinfo{journal}{Journal of the Physical Society of Japan}}
  \textbf{\bibinfo{volume}{88}}, \bibinfo{pages}{025001}
  (\bibinfo{year}{2019}).

\bibitem{cardy1996scaling}
\bibinfo{author}{Cardy, J.}
\newblock \emph{\bibinfo{title}{Scaling and renormalization in statistical
  physics}}, vol.~\bibinfo{volume}{5} (\bibinfo{publisher}{Cambridge Univ.
  Press}, \bibinfo{year}{1996}).

\bibitem{goldenfeld2018lectures}
\bibinfo{author}{Goldenfeld, N.}
\newblock \emph{\bibinfo{title}{Lectures on phase transitions and the
  renormalization group}} (\bibinfo{publisher}{CRC Press},
  \bibinfo{year}{2018}).

\bibitem{Landau}
\bibinfo{author}{Landau, L.} \& \bibinfo{author}{Lifshitz, E.}
\newblock \emph{\bibinfo{title}{Elasticity theory}}
  (\bibinfo{publisher}{Pergamon Press}, \bibinfo{year}{1975}).

\bibitem{SM1}
\bibinfo{title}{See supplemental material at [url will be inserted by
  publisher] for further details.} .

\bibitem{rafsanjani2019propagation}
\bibinfo{author}{Rafsanjani, A.}, \bibinfo{author}{Jin, L.},
  \bibinfo{author}{Deng, B.} \& \bibinfo{author}{Bertoldi, K.}
\newblock \bibinfo{title}{Propagation of pop ups in kirigami shells}.
\newblock \emph{\bibinfo{journal}{Proceedings of the National Academy of
  Sciences}} \textbf{\bibinfo{volume}{116}}, \bibinfo{pages}{8200--8205}
  (\bibinfo{year}{2019}).

\bibitem{moshe2019nonlinear}
\bibinfo{author}{Moshe, M.} \emph{et~al.}
\newblock \bibinfo{title}{Nonlinear mechanics of thin frames}.
\newblock \emph{\bibinfo{journal}{Physical Review E}}
  \textbf{\bibinfo{volume}{99}}, \bibinfo{pages}{013002}
  (\bibinfo{year}{2019}).

\bibitem{moshe2019kirigami}
\bibinfo{author}{Moshe, M.} \emph{et~al.}
\newblock \bibinfo{title}{Kirigami mechanics as stress relief by elastic
  charges}.
\newblock \emph{\bibinfo{journal}{Physical review letters}}
  \textbf{\bibinfo{volume}{122}}, \bibinfo{pages}{048001}
  (\bibinfo{year}{2019}).

\bibitem{yang2018multistable}
\bibinfo{author}{Yang, Y.}, \bibinfo{author}{Dias, M.~A.} \&
  \bibinfo{author}{Holmes, D.~P.}
\newblock \bibinfo{title}{Multistable kirigami for tunable architected
  materials}.
\newblock \emph{\bibinfo{journal}{arXiv preprint arXiv:1807.06498}}
  (\bibinfo{year}{2018}).

\bibitem{pippard1985response}
\bibinfo{author}{Pippard, A.~B.}
\newblock \emph{\bibinfo{title}{Response and stability: An introduction to the
  physical theory}} (\bibinfo{publisher}{CUP Archive}, \bibinfo{year}{1985}).

\bibitem{audoly2010elasticity}
\bibinfo{author}{Audoly, B.} \& \bibinfo{author}{Pomeau, Y.}
\newblock \emph{\bibinfo{title}{Elasticity and Geometry: From Hair Curls to the
  Non-linear Response of Shells}} (\bibinfo{publisher}{Oxford University
  Press}, \bibinfo{year}{2010}).

\bibitem{cedolin2010stability}
\bibinfo{author}{Cedolin, L.} \emph{et~al.}
\newblock \emph{\bibinfo{title}{Stability of structures: elastic, inelastic,
  fracture and damage theories}} (\bibinfo{publisher}{World Scientific},
  \bibinfo{year}{2010}).

\bibitem{bobnar2011euler}
\bibinfo{author}{Bobnar, J.} \emph{et~al.}
\newblock \bibinfo{title}{Euler strut: a mechanical analogy for dynamics in the
  vicinity of a critical point}.
\newblock \emph{\bibinfo{journal}{European Journal of Physics}}
  \textbf{\bibinfo{volume}{32}}, \bibinfo{pages}{1007} (\bibinfo{year}{2011}).

\bibitem{dias2017kirigami}
\bibinfo{author}{Dias, M.~A.} \emph{et~al.}
\newblock \bibinfo{title}{Kirigami actuators}.
\newblock \emph{\bibinfo{journal}{Soft matter}} \textbf{\bibinfo{volume}{13}},
  \bibinfo{pages}{9087--9092} (\bibinfo{year}{2017}).

\bibitem{livi2017nonequilibrium}
\bibinfo{author}{Livi, R.} \& \bibinfo{author}{Politi, P.}
\newblock \emph{\bibinfo{title}{Nonequilibrium statistical physics: a modern
  perspective}} (\bibinfo{publisher}{Cambridge University Press},
  \bibinfo{year}{2017}).

\end{thebibliography}
\newpage

\section*{Supplemental Material}

\subsection{Bending energy of a plate and the effect of stretching}

\label{A3}

\begin{figure}[h]
\includegraphics[width=0.3\textwidth]{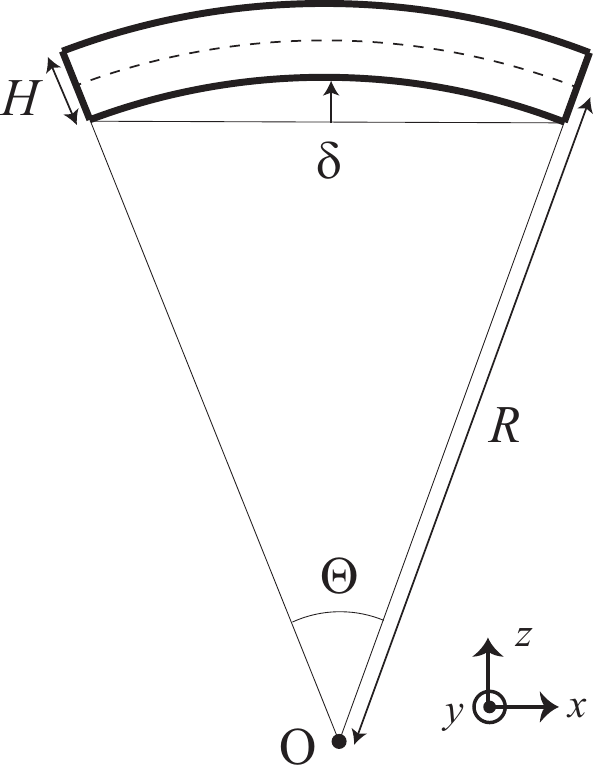}\caption{Bending of a plate of
thickness $H$, length $L$, and width $W$ (in the $y$ direction).}%
\label{FigA}%
\end{figure}

We consider that a plate of thickness $H$ (in the $z$ direction in the
undeformed plane state), length $L$ (in the $x$ direction in the plane state),
and width $W$ (in the $y$ direction) bends such that the sides of length $L$
of the plate become an arc of radius $R$ where the arc and its center O are
located on the $x-z$ plane (see Fig. \ref{FigA}). If we consider $r-\theta$
coordinate around this center on the $x-z$ plane, the plate occupies the
region defined by the conditions $-H/2+R<r<R+H/2$ and $-\Theta/2<\theta
<\Theta/2$. When "net stretch" is absent, the area of the "middle surface"
located at $r=R$ (represented by the dashed line in Fig. \ref{FigA}) remains
the original value $LW$. When "net stretch" comes into play, this area becomes
$L(1+\varepsilon_{0})W$ with $\varepsilon_{0}$ describing the size of stretch
in the $\theta$ direction, whereas%
\begin{equation}
R\Theta=L(1+\varepsilon_{0}) \label{A10}%
\end{equation}
The stretch associated with the surface located as $r=R+\eta$ is given by
$L(\eta)=(R+\eta)\Theta$ and the strain in the $\theta$ direction can be
estimated as $\varepsilon(\eta)=(L(\eta)-L)/L$, which gives%
\begin{equation}
\varepsilon(\eta)=\varepsilon_{0}+\eta(1+\varepsilon_{0})/R\simeq
\varepsilon_{0}+\eta/R
\end{equation}
when $\varepsilon_{0}\ll1$. The bending energy thus scales as%
\begin{equation}
U\simeq\int_{-H/2}^{H/2}d\eta E\varepsilon(\eta)^{2}WL(\eta)/2,
\end{equation}
in which $\varepsilon(\eta)^{2}L(\eta)\simeq(\eta/R+\varepsilon_{0})^{2}L$ for
$\varepsilon_{0}\ll1$ and $\eta\ll R$. (The proportional coefficient is in
general dependent on Poison's ratio and boundary condition. Here, we focus on
relation at the level of scaling laws and ignore the coefficient.) As a
result, we obtain
\begin{equation}
U\simeq\frac{EWLR}{6}\left[  \left(  \varepsilon_{0}+\frac{H}{2R}\right)
^{3}-\left(  \varepsilon_{0}-\frac{H}{2R}\right)  ^{3}\right]
\end{equation}
which gives%
\begin{equation}
U\simeq\frac{EWLR}{3}\left[  \left(  \frac{H}{2R}\right)  ^{3}+3\frac{H}%
{2R}\varepsilon_{0}^{2}\right]  \label{A11}%
\end{equation}

Introducing $\delta$ as the size of bending in the $z$ direction (see Fig.
\ref{FigA}) and considering the geometrical relations, $R^{2}=(R-\delta
)^{2}+(L(1+\varepsilon_{0})\cos(\Theta/2)/2)^{2}$ and $\delta=(L/2)\tan
(\Theta/2)$ with Eq. (\ref{A10}), we obtain
\begin{align}
2\delta R  &  =L^{2}/4\label{A12}\\
\Theta &  =4\delta/L \label{A12B}%
\end{align}
for $\varepsilon_{0}\ll1$ and $L\ll R$.

Equation (\ref{A12}) reduces the first term in Eq. (\ref{A11}) to the standard
form $\simeq(8/3)EWLH^{3}(\delta/L^{2})^{2}$, from which we can derive Eqs.
(\ref{U1}) and (\ref{U2}), at the level of scaling laws: the in-plane and
out-of-plane deformations corresponding to $(W,L,H,\delta)=(h,w,d,\delta
_{\parallel})$ and $(W,L,H,\delta)=(d,w,h,\delta_{\perp})$, respectively. The
correction term in Eq. (\ref{A11}) proportional to $\varepsilon_{0}^{2}$
always produces a positive term, which means any "net stretch" associated with
bending is energetically unfavorable.

Stretching can nevertheless occur in some cases, depending on the boundary
condition \cite{Landau}. For example, when one tries to bend a circular plate,
if the length of the circular edge is fixed the deformation is accompanied
with stretch of the diameter (here, we consider a circle on a spherical
surface); on the other hand, if the length of the diameter is fixed, the
length of the circular edge should be shrunk.

In the present example, we can set the following boundary constraint (the
length between the both ends of the plate fixed to $L$):%
\begin{equation}
L=L(1+\varepsilon_{0})\cos(\Theta/2) \label{A13}%
\end{equation}
For $\varepsilon_{0}\ll1$ and $L\ll R$, this boundary condition results in%
\begin{equation}
\varepsilon_{0}=(\Theta/2)^{2}/2=2(\delta/L)^{2}%
\end{equation}
Under this edge boundary constraint set by Eq. (\ref{A13}), the full energy
given in Eq. (\ref{A11}) can be expressed as%

\begin{equation}
U\simeq EWL\left[  (8/3)H^{3}(\delta/L^{2})^{2}+2H(\delta/L)^{4}\right]
\label{A14}%
\end{equation}
The bending and stretch energies scale as $EWLH^{3}(\delta/L^{2})^{2}$ and
$EWLH(\delta/L)^{4}$, respectively, and the ratio is $H^{2}:\delta^{2}$. The
same scaling laws are given in \cite{Landau} in a more general context.

However, the bending deformation of a unit element of the present kirigami
shown in Fig. \ref{f1}(a) occurs under no length constraint in the direction
of $w$ (corresponding to the direction of $L$ in the present example). In the
case of our kirigami in Fig.\ref{f1}(a), the ($2n-1$)-th [2$n$-th] unit
element is constraint at the middle section (of length $d$) of the top
[bottom] side of length $w+2d$, where the unit is connected to ($2n-2$)-th
[($2n-1$)-th] unit. However, the left and right edge sections (of length $d$)
of the bottom [top] side (of length $w+2d$) of the ($2n-1$)-th [$2n$-th] unit,
where the unit is connected to the 2$n$-th [(2$n-1$)-th] unit, are not
constrained and can freely move in the direction of $w$. [As a result, to
minimize the energy, the distance between both edges of the unit element tend
to become closer (because the edges are not constrained) as the kirigami
sample is stretched at the top and bottoms ends.] Thus, stretching is not
expected play any dominant roles in the present kirigami.

\subsection{Analytical details of the present model}

\label{A2}

We consider the expansion of $\tilde{U}(\tilde{\delta},\theta)$ in terms of
$\theta$%
\begin{equation}
\tilde{U}(\tilde{\delta},\theta)=\tilde{U}(\tilde{\delta},0)+A\theta
^{2}+b\theta^{4}+c\theta^{6}+\cdots\label{AA1}%
\end{equation}
and examine the behavior of $\tilde{U}(\tilde{\delta},\theta)$ for small
$\theta$. For this purpose, we obtain the following expansion for the two
energies in Eq. (\ref{U12c}):%
\begin{align}
\tilde{U}_{\parallel}  &  =\tilde{\delta}^{2}-\tilde{\delta}(1+\tilde{\delta
})\theta^{2}+(\tilde{\delta}+3(1+\tilde{\delta}))(1+\tilde{\delta})\theta
^{4}/12+\cdots\\
\tilde{U}_{\perp}  &  =\tilde{h}^{2}(1+\tilde{\delta})^{2}(\theta^{2}%
-\theta^{4}/3+\cdots)
\end{align}
This implies%
\begin{align}
A  &  =(1+\tilde{\delta})(-\tilde{\delta}+(1+\tilde{\delta})\tilde{h}^{2})\\
b  &  =(1+\tilde{\delta})((3+4\tilde{\delta})-4(1+\tilde{\delta})\tilde{h}%
^{2})/12,
\end{align}
which can be re-expressed as in the following form (note that the following
quantities can be obtained by the relation $A=[\partial^{2}\tilde{U}%
(\tilde{\delta},\theta)/\partial\theta^{2}]_{\theta=0}/2$ and $b=[\partial
^{4}\tilde{U}(\tilde{\delta},\theta)/\partial\theta^{4}]_{\theta=0}/4!$):%
\begin{align}
A  &  =-(1-\tilde{h}^{2})(\tilde{\delta}+1)(\tilde{\delta}-\tilde{\delta}%
_{c})\\
b  &  =(1/3)(1-\tilde{h}^{2})(\tilde{\delta}+1)(\tilde{\delta}+\tilde{\delta
}_{4})
\end{align}
with $\tilde{\delta}_{c}$ given by $\tilde{h}^{2}/(1-\tilde{h}^{2})$ as in Eq.
(\ref{delc}) and with $\tilde{\delta}_{4}$ given by%
\begin{equation}
4\tilde{\delta}_{4}=(3-4\tilde{h}^{2})/(1-\tilde{h}^{2}),
\end{equation}
which suggest that, for $\tilde{h}<1$ and $\tilde{\delta}>0$ as in the present
case, the coefficient $b$ is positive and the coefficient $A$ changes the sign
at $\tilde{\delta}=\tilde{\delta}_{c}$. In such a case, for small $\theta$,
Landau's continuous transition is expected: a continuous transition occurs at
$\tilde{\delta}=\tilde{\delta}_{c}\,$, i.e., when $A=0$. This is because, for
the values $\tilde{\delta}$ at which $A$ is positive, $\tilde{U}(\tilde
{\delta},\theta)$ as a function of $\theta$ has a single minimum at $\theta=0$
(note that $b$ is positive), but for the values $\tilde{\delta}$ at which $A$
is negative, $\tilde{U}(\tilde{\delta},\theta)$ as a function of $\theta$ has
double minima at $\theta=\pm\theta^{\ast}$. The value of $\theta^{\ast}$ is
obtained by solving the equation $\partial\tilde{U}(\tilde{\delta}%
,\theta)/\partial\theta=0$ in terms of $\theta$. From this equation, we obtain
$\theta^{\ast}=0$ and, for $\tilde{\delta}>\tilde{\delta}_{c}$, another
solution $(\theta^{\ast})^{2}=-A/(2b)=(3/2)(\tilde{\delta}-\tilde{\delta}%
_{c})/(\tilde{\delta}+\tilde{\delta}_{4})$, which behaves as $2(1-\tilde
{h}^{2})(\tilde{\delta}-\tilde{\delta}_{c})$ in the vicinity of $\tilde
{\delta}=\tilde{\delta}_{c}$. In this way, we can derive Eq. (\ref{theta}) and
we can justify analytically that the point $\delta=\delta_{c}$ is truly the
transition point. From Eqs. (\ref{AA1}), (\ref{U12c}) and (\ref{theta}), for
$\tilde{\delta}$, $\theta^{\ast}$, $\tilde{h}\ll1$, we obtain
\begin{align}
\tilde{U}(\tilde{\delta},\theta^{\ast})  &  =\left\{
\begin{array}
[c]{cc}%
\tilde{\delta}^{2}=\tilde{\delta}_{c}^{2}(\delta/\delta_{c})^{2} & \text{for
}\delta<\delta_{c}\\
\tilde{\delta}_{c}^{2}+2\tilde{h}^{2}(\tilde{\delta}-\tilde{\delta}_{c}) &
\text{for }\delta>\delta_{c}%
\end{array}
\right.  ,\label{Um}\\
\tilde{F}(\tilde{\delta})  &  =\frac{\partial\tilde{U}(\tilde{\delta}%
,\theta^{\ast})}{\partial\tilde{\delta}}=\left\{
\begin{array}
[c]{cc}%
2\tilde{\delta} & \text{for }\delta<\delta_{c}\\
2\tilde{h}^{2}=2\tilde{\delta}_{c} & \text{for }\delta>\delta_{c}%
\end{array}
\right.  . \label{F}%
\end{align}
In Fig. \ref{f4}(a), (b), and (c), the plots labeled as "Present Model
(approx.)" are based on Eqs. (\ref{theta}), (\ref{Um}), and (\ref{F}), respectively.

\subsection{Analytical details of the previous model}

\label{A1}

In our previous work, we simply considered two possible modes of deformations
and \textit{forbid} the mixture of the two modes: (1) "purely in-plane
deformation," in which $(\delta_{\perp},\theta)=(0,0)$, i.e., $\vec{\delta
}=\delta_{\Vert}\hat{y}$ together with $\delta=\delta_{\Vert}$, and (2)
"purely out-of plane deformation," in which $\delta_{\Vert}=0$ with a finite
angle $\theta=\theta^{\ast}$, i.e., $\vec{\delta}=\delta_{\perp}\hat{n}$
together with $\tan\theta^{\ast}=\delta_{\perp}/d$ and $(d+\delta)^{2}%
=d^{2}+\delta_{\perp}^{2}$, which lead to%
\begin{equation}
\theta^{\ast}=\left\{
\begin{array}
[c]{cc}%
0 & \text{for }\delta<2\delta_{c}\\
\arctan\left(  (1+\tilde{\delta})^{2}-1\right)  ^{1/2} & \text{for }%
\delta>2\delta_{c}%
\end{array}
\right.  \label{A21}%
\end{equation}
The energy for the former deformation scales with $\delta^{2}$ (because of Eq.
(\ref{U}) with $\delta_{\perp}=0$ and $\delta_{\Vert}=\delta$) and that for
the latter scales with $\delta$ (because of Eq. (\ref{U}) with $\delta_{\Vert
}=0$, together with $\delta_{\perp}^{2}=2d\delta$ in the limit $\delta\ll d$).
This means that at lower energies the purely in-plane deformation has the
lower energy but at higher energies the purely out-of-plane deformation
becomes the lower. The transition point for $\delta$ is determined by matching
the two energies, which is revealed to occur at $\delta=2\delta_{c}$. In
summary, the previous theory results in the following expressions in the limit
$\tilde{h}\ll1$:%
\begin{align}
\tilde{U}(\tilde{\delta})  &  =\left\{
\begin{array}
[c]{cc}%
\tilde{\delta}^{2} & \text{for }\delta<2\delta_{c}\\
2\tilde{h}^{2}\tilde{\delta}=2\tilde{\delta}_{c}\tilde{\delta} & \text{for
}\delta>2\delta_{c}%
\end{array}
\right. \label{Um1}\\
\tilde{F}(\tilde{\delta})  &  =\left\{
\begin{array}
[c]{cc}%
2\tilde{\delta} & \text{for }\delta<2\delta_{c}\\
2\tilde{h}^{2}=2\tilde{\delta}_{c} & \text{for }\delta>2\delta_{c}%
\end{array}
\right.  , \label{F2}%
\end{align}
In Fig. \ref{f4}(a), (b), and (c), the plots labeled as "Previous Model" are
based on Eqs. (\ref{A21}), (\ref{Um1}), and (\ref{F2}), respectively.

\subsection{Buckling of thin plates and Landau theory of critical phenomena}

\label{A5}

We showed in the main text that kirigami's transition can be viewed as the
critical phenomenon described by Landau theory. A similar scenario emerges for
the classic problem of buckling of thin plates (or beams) as explained in the following.

We again consider the situation in Fig. \ref{FigA}. However, this time we
assume that the distance between the left and right ends becomes $L-\Delta$
from the original value $L$ with $\Delta>0$. In the case of pure bending, the
area of the "middle surface" located at $r=R$ (represented by the dashed line
in Fig. \ref{FigA}) remains the original value $LW$. In general, this area
becomes $L(1+\varepsilon_{0})W$ with $\varepsilon_{0}(<0)$ describing the size
of shrinkage of the length of the neutral line in the $\theta$ direction to
obtain Eq. (\ref{A11}) as before.

Introducing $\delta$ as the size of bending in the $z$ direction (see Fig.
\ref{FigA}) as before and considering the geometrical relations,
$R^{2}=(R-\delta)^{2}+(L(1+\varepsilon_{0})\cos(\Theta/2)/2)^{2}$ and
$\delta=((L-\Delta)/2)\tan(\Theta/2)$ with Eq. (\ref{A10}), we again obtain
Eqs. (\ref{A12}) and (\ref{A12B}) for $\left\vert \varepsilon_{0}\right\vert
\ll1$ and $\Delta\ll L\ll R$.

In the present example, we have set the following boundary constraint (the
length between the both ends of the plate fixed to $L-\Delta$):%
\begin{equation}
L-\Delta=L(1+\varepsilon_{0})\cos(\Theta/2) \label{A13d}%
\end{equation}
From this, we decompose $\Delta$ in the two components:%
\begin{equation}
\Delta=\Delta_{B}+\Delta_{C} \label{delBC}%
\end{equation}
with introducing the pure bending part and pure compression part defined,
respectively, as
\begin{align}
\Delta_{B}  &  =L(\Theta/2)^{2}/2=2\delta^{2}/L\label{delB}\\
\Delta_{C}  &  =\left\vert \varepsilon_{0}\right\vert L \label{delC}%
\end{align}
Under the edge boundary constraint set by Eq. (\ref{A13d}), the full energy
given in Eq. (\ref{A11}) can be expressed as%

\begin{equation}
U\simeq\frac{4EWH^{3}}{3L^{2}}\Delta_{B}+\frac{EWH}{2L}\Delta_{C}^{2}%
\end{equation}
The first term corresponds to the pure bending deformation scaling as
$EH^{3}(\delta/L^{2})^{2}WL$. The second term to the pure compression, in
which the plate remains flat and is subject to a homogeneous strain of
$\varepsilon=\Delta/L$: the energy for pure compression scales as
$E\varepsilon^{2}HWL=EWH\Delta^{2}/L$, which coincides with the second term at
the level of scaling laws.

If we disallow the simultaneous existence of bending and compression, and
compare the two energies, $(4/3)EWH^{3}\Delta/L^{2}$ and $(1/2)EWH\Delta
^{2}/L$, which, respectively scale as $\Delta$ and $\Delta^{2}$, the pure
compression energy $\sim\Delta^{2}$ is lower than the pure bending energy
$\sim\Delta$ for small $\Delta$: the pure compression without bending is
expected for small $\Delta$. However, the relative importance of the two
energies is interchanged at the matching point: $\Delta=2\Delta_{c}$ with
\begin{equation}
\Delta_{c}=(4/3)H^{2}/L. \label{Adelc}%
\end{equation}
For large $\Delta$ ($\Delta>2\Delta_{c}$), the pure bending energy is lower
than the pure compression energy and, thus, the pure bending is predicted for
large $\Delta$. At the critical strain $\varepsilon=2\varepsilon_{c}$ with
$\varepsilon_{c}=\Delta_{c}/L=(4/3)(H/L)^{2}$, the force acting on the both
edges is derived by differentiating the bending energy with respect to
$\Delta$ to obtain a result for the critical force divided by the width $W$:
$P_{c}=F_{c}/W=(4/3)EH^{3}/L^{2}$. This corresponds to Euler buckling load
$\pi^{2}EI/L^{2}$, where $I\simeq H^{3}$ in the present case. If we derive the
force from the compression energy evaluated at the transition point
$\Delta=2\Delta_{c}$, we obtain instead $(8/3)EH^{3}/L^{2}$, i.e., twice the
value obtained from the bending energy. This difference between the forces at
the transition point corresponds to the force jump in the kirigami model.

If we allow the simultaneous existence of bending and compression (this
corresponds to the usual treatment in the field), we have to consider the full
energy given in Eq. (\ref{A14}). For convenience, we change the set of unknown
variables from $(\Delta_{B}$, $\Delta_{C})$ to $(\Delta$, $\delta)$ to have
the following expression with the aid of Eqs. (\ref{delBC}) to (\ref{delC}):%
\begin{equation}
U\simeq\frac{8EWH^{3}}{3L^{3}}\delta^{2}+\frac{2EWH}{L^{3}}\left(  \delta
^{2}-L\Delta/2\right)  ^{2} \label{As1}%
\end{equation}
We renormalize the energy and length scales, respectively, by the energy and
length scales $2EW^{5}H/L^{3}$ and $W$:%
\begin{equation}
\tilde{U}\simeq-\tilde{L}(\tilde{\Delta}-\tilde{\Delta}_{c})\tilde{\delta}%
^{2}+\tilde{\delta}^{4}+\left(  \tilde{L}\tilde{\Delta}/2\right)  ^{2}%
\end{equation}
with $\tilde{\Delta}_{c}$ defined in Eq. (\ref{Adelc}).

The last form is a typical Landau free energy, if we identify $\tilde{\delta}$
as the order parameter and $\tilde{\Delta}$ as the inverse temperature. When
the compression is small ($\tilde{\Delta}<\tilde{\Delta}_{c}$), the
coefficient of $\tilde{\delta}^{2}$ is positive: $\tilde{U}$ has a single
minimum at $\tilde{\delta}=0$, which corresponds to the pure compression
without bending. When the compression is large ($\tilde{\Delta}>\tilde{\Delta
}_{c}$), the coefficient becomes negative and thus $\tilde{U}$ as a function
of $\tilde{\delta}$ is convex near $\tilde{\delta}=0$; however, because of the
coefficient of $\tilde{\delta}^{4}$ is positive, we have double minima at
$\tilde{\delta}=\pm\tilde{\delta}^{\ast}$. This $\tilde{\delta}^{\ast}$ is the
theoretically predicted value of $\tilde{\delta}$ for $\tilde{\Delta}%
>\tilde{\Delta}_{c}$ and is determined as one of the solutions of
$\partial\tilde{U}/\partial\tilde{\delta}=0$, i.e., $-2\tilde{L}(\tilde
{\Delta}-\tilde{\Delta}_{c})\tilde{\delta}+4\tilde{\delta}^{3}=0$.
\begin{equation}
\tilde{\delta}^{\ast}=\left\{
\begin{array}
[c]{cc}%
0 & \text{ for }\tilde{\Delta}<\tilde{\Delta}_{c}\\
\lbrack\tilde{L}(\tilde{\Delta}-\tilde{\Delta}_{c})/2]^{1/2} & \text{for
}\tilde{\Delta}>\tilde{\Delta}_{c}%
\end{array}
\right.  \text{ }%
\end{equation}
This theory predicts a continuous transition for $\tilde{\delta}$ with the
classic exponent 1/2. The counterpart of Euler buckling load is given from
$F=\partial U/\partial\Delta$ with Eq. (\ref{As1}) evaluated at $\Delta
=\Delta_{c}$: $P_{c}=F_{c}/W=(4/3)EH^{3}/L^{2}$.

\end{document}